# Nonlocal gap solitons in parity-time symmetric optical lattices

Huagang Li,[1] Xiujuan Jiang,[2] Xing Zhu[3] and Zhiwei Shi,[2,*]

[1] *Department of Physics, Guangdong university of Education , Guangzhou 510303, P.R.China*
[2] *School of Information Engineering, Guangdong University of Technology, Guangzhou 510006, P.R.China*
[3] *State Key Laboratory of Optoelectronic Materials and Technologies, Sun Yat-Sen University, Guangzhou 510275, P.R.China*
[*] *Corresponding author: szwstar@gdut.edu.cn*

We numerically study the nonlocal gap solitons in parity-time (PT) symmetric optical lattices built into a nonlocal self-focusing medium. We state the existence, stability, and propagation dynamics of such PT gap solitons in detail. Simulated results show that there exist stable gap soltions. The influences of the degree of nonlocality on the soliton power, the energy flow density and the stable region of the PT gap solitons are also examined. © 2011 Optical Society of America

OCIS codes: 190.3270, 190.6135.

Self-action of light in periodic photonic crystals with periodic modulation of the refractive index generates rich optical phenomena [1]. The photonic crystals can provide efficient control of the transmission and localization of light, and they open the ways to tailor the diffraction and the route of the electromagnetic waves [2]. Quite a lot of nonlinear optical phenomena in nonlinear photonic crystals have been reported, including the self-trapping of localized modes in the form of gap solitons [3] which was also studied in other materials, layered microstructures, fiber Bragg gratings, Bose-Einstein condensates, waveguide arrays, and optically induced lattices. Nonlocal effects come to play an important role as the characteristic correlation radius of the medium's response function becomes comparable to the transverse width of the wave packet [4]. Nonlocality of nonlinear response may drastically modify the conditions necessary for gap soliton existence [1].

Recently, the solitons in synthetic optical media with parity-time (PT) symmetries have caught much attention [5–14,17–23]. Musslimani and the cooperator firstly discovered that a novel class of nonlinear self-trapped modes exist in optical PT synthetic lattices [5] and PT periodic structures exhibit unique characteristics stemming from the nonorthogonality of the associated Floquet-Bloch modes [6]. The behavior of a PT optical coupled system judiciously involving a complex index potential was observed in the experiment in 2010 [7, 8]. It was also stated the analytical solutions to a class of nonlinear Schrödinger equations with PT-like potentials [9], the stable dissipative defect modes in both focusing and defocusing media where periodic optical lattices were imprinted in the cubic nonlinear media with strong two-photon absorption [10], and the defect solitons in parity-time periodic potentials [11] . We also reported the gray solitons in PT symmetric potentials [12] and the gap solitons in PT complex periodic optical lattices with the real part of superlattices [13]. However, thus far all studies focus on the local nonlinear media with the PT symmetry potentials, and the solitons supported by the nonlocal nonlinear media with the PT symmetry optical lattices are never reported.

In this paper, the gap solitons in the PT symmetric optical lattices built into a nonlocal self-focusing medium are studied. We state the existence, stability, and propagation dynamics of such PT gap solitons in detail. Simulated results show that there exist stable gap solitons. In addition, we find that the degree of nonlocality can influence the soliton power, the energy flow density, and the region where the stable PT gap solitons can exist.

In a nonlocal self-focusing medium with PT symmetric optical lattice, the one-dimensional optical wave propagation can be described by the normalized nonlinear Schrödinger (NLS)-like equation

$$i\frac{\partial q}{\partial z} + \frac{\partial^2 q}{\partial x^2} + Rq + q\int_{-\infty}^{+\infty} g(x-\lambda)|q(\lambda)|^2 d\lambda = 0, \quad (1)$$

where $q$ is the complex dimensionless light field amplitude, $z$ is the normalized longitudinal coordinate, and $g$ is the nonlocal response function. $R = V(x) + iW(x)$ is the PT symmetric optical lattice, and $V(x)$ and $W(x)$ are its real and imaginary components, respectively. We are going to search for a stationary soliton solution of Eq. (1) in the form of $q(x,z) = u(x)e^{ibz}$, where $u$ is a complex function and $b$ is the propagation constant of spatial solitons [15]. Thus, Eq. (1) can be changed into

$$\frac{\partial^2 u}{\partial x^2} + Ru - bu + u\int_{-\infty}^{+\infty} g(x-\lambda)|u(\lambda)|^2 d\lambda = 0. \quad (2)$$

Here, the nonlocality of the materials is supposed to be ruled with an exponential response function $g(x) = 1/(2d^{1/2})\exp(-|x|/d^{1/2})$ (as in liquid crystals), where $d$ is the degree of the nonlocality. We assume a PT symmetric optical lattices in which $V(x) = V_0\{\cos[2\pi\sin(x)]+1\}/2$ and $W(x) = W_0\sin(x)$, and $V_0$ and $W_0$ are the amplitudes of the real and imaginary parts. Although the PT symmetric optical lattice has crossed the phase transition point, the solitons still exist because the amplitude of the refractive index distribution can be altered by the beam itself through the optical nonlocal nonlinearity. The parity-time symmetry will remain broken if it cannot be nonlocal nonlinearly restored [5].

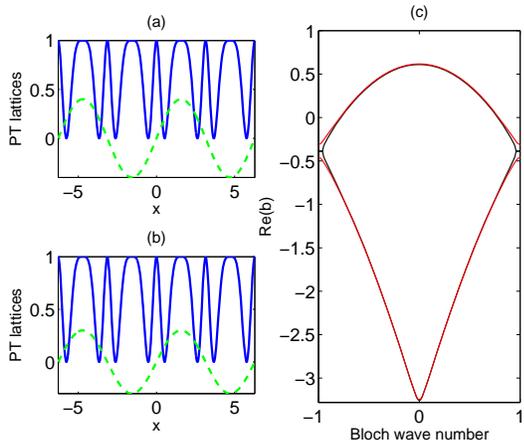

Fig. 1. (Color online) (a) PT complex periodic optical lattices when $V_0 = 1$, $W_0 = 0.4$. (b) PT complex periodic optical lattices when $V_0 = 1$, $W_0 = 0.3$. The solid blue and dash green curves represent the real part and the imaginary part of the optical lattices, respectively. (c) The band structures corresponding to (a) (the black curve) and (b) (the red curve).

The linear version of Eq.(2) is

$$\frac{\partial^2 u}{\partial x^2} + Ru - bu = 0, \qquad (3)$$

where $b$ now represents the propagation constant in the PT symmetric optical lattices. The Bloch theorem tells us that the eigenfunctions of Eq.(3) are in the form of $u = F_k exp(ikx)$, where $k$ is the Bloch wave number, and $F_k$ is a periodic function of $x$ with the same period as the lattices $R$. Substituting the Bloch solution into Eq.(3), we can get the eigenvalue equation and then obtain the band structure using the plane wave expansion method numerically [5, 12]. The PT symmetric optical lattices are shown in Fig. 1 (a) ($V_0 = 1, W_0 = 0.4$) and (b) ($V_0 = 1, W_0 = 0.3$). We numerically find that the purely real bands are possible in the range $0 \le W_0 \le 0.35$, and the region of the semi-infinite gap is $b \ge 0.62$ when $W_0 = 0.3$ [5]. In Fig. 1(c) we depict the associated band structures for various values of the potential parameter $W_0$ (below and above the phase transition point $W_0 = 0.35$). It is noticed that the band gap becomes narrower as $W_0$ increases, and the gap closes completely when crossing the critical transition value $W_0 = 0.35$ [5].

Based on the band-gap structure, we get the soliton solutions by solving Eq.(2) numerically using the modified squared-operator method [16]. We find a family of localized solutions with real eigenvalues located within the semi-infinite gap ($b \ge 0.62$). The typical cases of these gap solitons are shown in Fig. 2, where it is depicted in (a)-(d) the fields of the unstable and stable gap soliton at $d = 0.5$ and $d = 2$, respectively. To shed more light on the properties of the stable solitons, we study the parameter $S = (i/2)(uu_x^* - u^*u_x)$ associated with the transverse power flow density or Poynting vector across the beam [5]. The transverse power flow density $S$ is shown in Fig. 2(e) and (f), where one can find that $S$ is positive everywhere, which implies that the energy always flows in one direction, i.e., from the gain region toward the loss region [5].

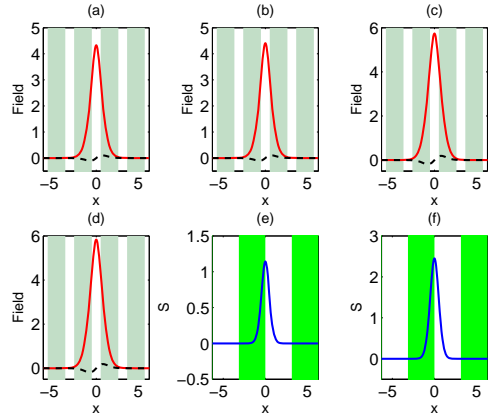

Fig. 2. (Color online) (a) and (b) Soliton profiles with $b = 7.48, 7.70$ at $d = 0.5$. (c) and (d) Soliton profiles with $b = 9.73, 9.98$ at $d = 2$. The solid red curves and the dash black curves represent the real part and the imaginary part, respectively. (e) and (f) Transverse power flow of the solitons shown in (b) and (d), respectively. The green regions represent the gain domains, and the white regions represent the loss domains. The potential parameters are $V_0 = 1$ and $W_0 = 0.3$.

To check the stability of the solitons with the method of linear stability analysis, we assume $q(x,z) = u(x)e^{ibz} + \epsilon[F(x)e^{i\delta z} + G^*(x)e^{-i\delta^* z}]e^{ibz}$, where $\epsilon \ll 1$, $F$ and $G$ are the perturbation eigenfunctions, and $\delta$ is the growth rate of the perturbation [5]. By linearizing Eq.(1), we gain

$$\delta F = \frac{\partial^2 F}{\partial x^2} + (V + iW)F + nF - bF + u\Delta n, \qquad (4)$$

$$\delta G = -\frac{\partial^2 G}{\partial x^2} + (-V + iW)G - nG + bG - u^*\Delta n, \qquad (5)$$

where $n = \int_{-\infty}^{+\infty} g(x-\lambda)|u(\lambda)|^2 d\lambda$, and $\Delta n = \int_{-\infty}^{+\infty} g(x-\lambda)[G(\lambda)u(\lambda) + F(\lambda)u^*(\lambda)]d\lambda$. The gap soltions are linearly unstable when $\delta$ has an imaginary component, on the contrary, they are stable when $\delta$ is real. In Fig. 3 (a), Im($\delta$), the imaginary component of $\delta$, is not always zero, so there are several stable regions for the gap solitons. Evidently, the degree of nonlocality can influence the stable region. The power of solitons is defined as $P = \int_{-\infty}^{+\infty} |u|^2 dx$, and $P$ is a monotonically-increasing function of the propagation constant $b$, as is shown in

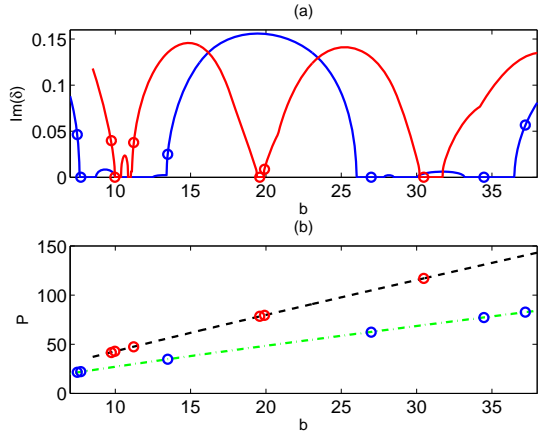

Fig. 3. (Color online) (a) Im($\delta$) versus $b$, the blue curve is for $d = 0.5$ and the red curve is for $d = 2$. (b) $P$ versus $b$, the dash-dotted green curve is for $d = 0.5$ and the dash black curve is for $d = 2$. The other parameters are $V_0 = 1$, $W_0 = 0.3$. Points marked with circle correspond to the cases shown in Fig. 4(a)-(l).

Fig. 3(b). To examine further the robustness of these PT lattice self-strapped modes, we also simulate the propagation of beams under different conditions, as presented in Fig. 4(a)-(f) for $d = 0.5$ and Fig. 4 (g)-(l) for $d = 2$, which support the above conclusion about the stability of nonlocal gap soltions as well.

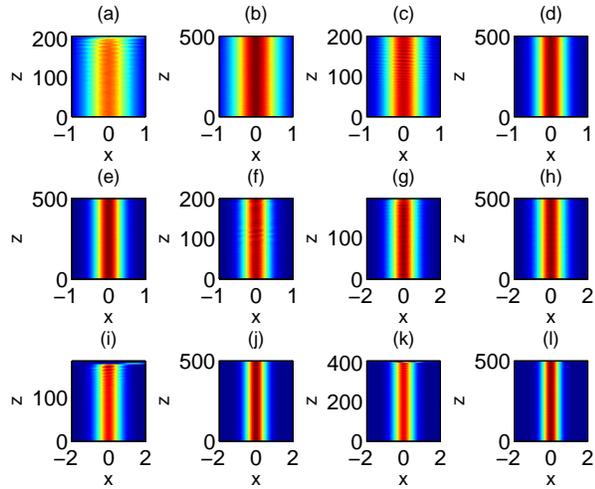

Fig. 4. (Color online) Simulated propagation of the solitons with 1% random noise. (a)-(f) $b = 7.48, 7.70, 13.48, 26.98, 34.48, 37.23$ at $d = 0.5$, respectively. (g)-(l) $b = 9.73, 9.98, 11.23, 19.60, 19.90, 30.48$ at $d = 2$, respectively. The other parameters are $V_0 = 1$, $W_0 = 0.3$.

In conclusion, we investigate the gap solitons in the PT symmetric optical lattices built into a nonlocal self-focusing medium. The existence, stability, and propagation dynamics of such PT gap solitons are stated in detail. Simulated results show that there exist stable gap soltions, and the degree of nonlocality can influence the soliton power, the energy flow density, and the region where stable PT gap solitons can exist.

This research is supported by the Natural Science Foundation of Guangdong Province, China (Grant No. 10151063101000017 ), the Open Fund of Key Laboratory for High Power Laser Physics of Chinese Academy of Science (Grant No. SG-001103), and the National Natural Science Foundation of China (Grant No. 10904041 and 10974061).

## Informational Fourth Page

In this section, please provide full versions of citations to assist reviewers and editors (OL publishes a short form of citations) or any other information that would aid the peer-review process.